**Astronomy**

**Starbursts near and far**

Yu Gao

**Observations of intensely bright star-forming galaxies both close by and in the distant Universe at first glance seem to emphasize their similarity. But look a little closer, and differences emerge.**

In a recent issue of the *Astrophysical Journal,* two papers[1,2] present the latest measurements of 'starburst' galaxies — galaxies whose extreme brightness is thought to indicate short, intense bursts of star formation. The latest work is exemplary of the instruments and methods that are providing ever deeper insight into these spectacular objects. But equally, it exposes gaps in our knowledge that will only be filled by the next generation of telescopes.

The first of the papers, by Mangum *et al.*[1], provides perhaps the most accurate measurements so far of the density of star-forming molecular gas in nearby starburst galaxies. The authors used the National Radio Astronomy Observatory's 100-metre-aperture Green Bank Telescope in West Virginia to survey radio wave emitted at frequencies corresponding to two so called *K*-doublet transitions of the organic molecule formaldehyde ($H_2CO$) - a reliable density and temperature probe in the star-forming molecular clouds in our own Galaxy - in 19 close, infrared-bright galaxies. Such measurements have proved unexpectedly complicated, even at such short ranges. One transition was detected from most of the galaxies, but only five could be monitored with detection in both transitions that allows precise measurement of the spatial densities. The transitions are simply rather weak to be both uncovered. And observations involve emission and/or absorption features that might sometimes be mingled and are thus problematic to robustly distinguish. Additionally, detailed modeling has to be revoked. Owing to the intricate excitation of the transitions, they appear in absorption for density smaller than about three hundred thousand hydrogen molecules per centimetre cube, whereas in emission above this threshold. The findings confirm what has been observed with other dense gas tracers such as hydrogen cyanide (HCN)[3] that galaxies with higher star formation rate tend to have higher gas density, supporting the idea that vigorous starbursts are driven by the amount of dense molecular gas available to form stars[4].

In the second study, Hathi *et al.*[2] extend observations of starburst intensities to much more distant galaxies at redshifts of up to 6 and compare with those, similar to Mangum and colleagues' nearby sample, at redshifts close to zero. Because the Universe is expanding, light from a distant cosmic source will be shifted to longer ('redder') wavelengths as it travels towards us. Redshift thus measures both 'lookback time' – light with a redshift of 6, for example, is light emitted when the Universe was about a tenth of its current age – and also, because light travels at a finite, constant speed through the cosmos, a source's distance from us.

New-born stars emit light mostly at ultraviolet wavelengths. This light heats the surrounding interstellar dust, which then radiates at infrared wavelengths. Comparing with far-infrared radiation, Mangum *et al.*[1] observe $H_2CO$ at much longer centimetre wavelength. Hathi *et al.*[2], by contrast, use two cameras aboard the Hubble Space Telescope sensitive at optical and near-infrared wavelengths to observe the original ultraviolet light from two galaxy samples, at redshifts of 3-4 and 5-6 respectively. They deduce constant maximum star formation intensity for both samples, one that is broadly in agreement with earlier observations of starbursts at lower redshifts.

But before we hasten to draw physical conclusions from this similarity, we must bear in mind that we might not be comparing like with like. The most powerful local starbursts, ultraluminous infrared galaxies or ULIRGs, typically have extreme starburst regions smaller than a thousand light years. The mass of gas in those regions is the equivalent of a billion solar masses, and the luminosity can be greater than that of a couple hundred billion Suns[5] - several times more than is emitted by an entire typical spiral galaxy such as our Milky Way. Local extreme starburst intensity, buried in the dust, seems to be already larger than the highest measured by Hathi *et al.*[2]

Ultraviolet imaging of the central regions of nearby ULIRGs indicates that ultraviolet emissions make up at most only around 7 per cent of total emissions[6], and that a thousand light years or more separate peaks of ultraviolet and infrared emission. The implication is that Hathi and colleagues' intensity measurements represent just residual light leaking from stars in dust-enshrouded star-forming regions, while missing the more dominant infrared radiation reradiated by the dust. That supposition is supported by recent surveys of the deep Universe, such as COSMOS, GOODS, SWIRE[7], which are finding more and more high-redshift dust-obscured galaxies with large infrared-to-ultraviolet luminosity ratios that had been missed in traditional optical surveys.

That would seem to lead to one of two conclusions. First, that Hathi and colleagues' high-redshift galaxies are powered by even brighter extreme starbursts mostly hidden in the dust as those found in the local Universe - with concomitantly much higher densities of molecular gas than could possibly be extrapolated from Mangum and colleagues' $H_2CO$ results. Recent efforts to detect HCN at high redshifts have offered some clue for a higher ratio of star formation rate to dense gas at early cosmic times[8,9,10]. Alternatively, the earlier starbursts might simply be very much larger in extent, with constant intensities comparable to the nearby starburst galaxies.

If the earlier starbursts are indeed much in general more intense, we might suppose they have a different physical origin. The extreme starburst activity of nearby ULIRGs is thought to have been triggered by the strong interaction or merger of gas-rich spiral galaxies (Fig. 1). A possibility for the high-redshift starbursts is that a fraction of the luminosity is caused by a dust-obscured 'active galactic nucleus' (AGN or more than one) – a black hole at the centre of the merging galaxies. If that is so, are there any evolutionary connections between extreme starbursts, the build-up of massive AGNs, and how galaxies assemble? A link between the black hole mass at the centre and the mass of the host galaxy bulge indeed exists[11,12]. Therefore, it is likely that central extreme starbursts help speed up the bulge mass assembly via rapid, out-of-control star formation, which is then closely linked with the growth of central black hole and hence higher black hole accretion rate - stronger AGN activity.

Combining optical and infrared lights is one way to improve our estimation of star formation rate[13]. But observations of distant starbursts at longer far-infrared and millimetre wavelengths will be the key to better quantify their true star formation rate and intensity. We shan't know the full answer until the resolving power and sensitivity of next-generation facilities such as the Atacama Large Millimeter/Submillimeter Array (ALMA) and NASA's James Webb Space Telescope come to fruition in the next few years[14]. ALMA will be able to reveal the obscured starburst intensity maps, as well as the distribution of dense molecular gas and its kinematics at high redshifts. The capability of JWST in the mid-infrared will directly expose hidden starbursts buried in their surrounding dust. We will have to wait a little longer, however, before we have the wherewithal to apply Mangum and colleagues' $H_2CO$ densitometer[1] to Hathi and colleagues' high-redshift galaxy sample[2]. But even that capability might come with the Square Kilometer Array, an international radio telescope currently at the planning stage.

Yu Gao is at Purple Mountain Observatory, Chinese Academy of Sciences, Nanjing, Jiangsu 210008, China.
e-mail: yugao@pmo.ac.cn

**Figure 1 Buried in dust.** These two images of the spectacular merger of the nearby Antennae galaxies - on the left from NASA's Spitzer Space Telescope, on the right from the Hubble Space Telescope - show the train wreck of two gas-rich spirals. The proximity of the galaxies allows detailed imaging of new-born star clusters and the infrared hotspots that mark starburst sites. The dusty regions between the two galaxies, heavily obscured in the HST optical image, are in fact the dominant sites of active star formation as traced by the Spitzer Infrared Array Camera at an infrared wavelength of 8 micrometre emission (red). Indeed, the brightest infrared hot spot (bottom left of the Spitzer image) is almost entirely unseen. Extreme starbursts near and far are more than several ten times brighter than the Antennae galaxies. And the most intense starbursts are usually hidden in dust, making comparisons between observations at different redshifts particularly tricky.

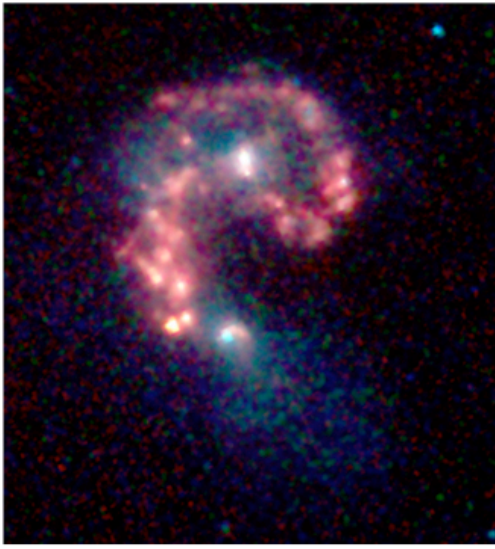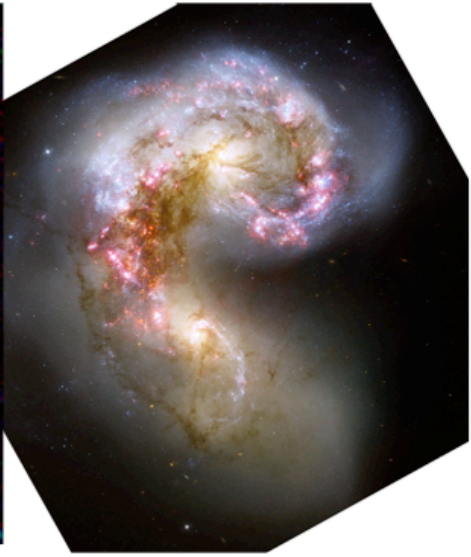